\documentclass[10pt,twocolumn,letterpaper]{IEEEtran}
\usepackage{psfig}
\usepackage{graphicx}
\usepackage{epsfig}
\usepackage{latexsym}
\usepackage{amsfonts}
\usepackage{here}
\usepackage{rawfonts}
\usepackage[latin1]{inputenc}
\usepackage[T1]{fontenc}
\usepackage{calc}
\usepackage{url}
\usepackage{enumerate}
\usepackage{color}
\usepackage{amssymb}
\usepackage{bm}
\usepackage{upref}
\usepackage{times}
\usepackage{amsmath}
\usepackage{cite}

\newcommand{\bs}{\mathbf}

\newcommand{\undb}{\bs}
\newcommand{\beq}{\begin{equation}}
\newcommand{\eeq}{\end{equation}}
\newcommand{\beqarr}{\begin{array}}
\newcommand{\eeqarr}{\end{array}}

\newtheorem{theorem}{{\bf Theorem}}

\newcommand{\qed}{\nobreak \ifvmode \relax \else
  \ifdim\lastskip<1.5em \hskip-\lastskip
  \hskip1.5em plus0em minus0.5em \fi \nobreak
  \vrule height0.75em width0.5em depth0.25em\fi}

\newcounter{step}
\newlength{\totlinewidth}
  {\end{list}%
  \rule{\linewidth}{1pt}}
\newcounter{substep}

  {\end{list}}

\newlength{\aligntop}
\newlength{\alignbot}
\makeatletter
\renewenvironment{align}{%
  \vspace{\aligntop}
  \start@align\@ne\st@rredfalse\m@ne
}{%
  \math@cr \black@\totwidth@
  \egroup
  \ifingather@
    \restorealignstate@
    \egroup
    \nonumber
    \ifnum0=`{\fi\iffalse}\fi
  \else
    $$%
  \fi
  \ignorespacesafterend%
  \vspace{\alignbot}\par\noindent
}
\makeatother

\IEEEoverridecommandlockouts

\author{
\medskip 
\normalsize 
{$\mbox{Arti M.K.}$, {\em Student Member, IEEE} and $\mbox{Manav R. Bhatnagar}$, {\em Senior Member, IEEE}  \vspace*{-2.0em}}
\thanks{Arti M.K. and Manav R. Bhatnagar are
with the Department of Electrical
Engineering, Indian Institute of Technology - Delhi, Hauz Khas,
New Delhi 110016, India (e-mails:
{\tt \{arti.mk,manav\}@ee.iitd.ac.in}).}
}\date{}
\renewcommand{\baselinestretch}{1.00}

\title{Performance Analysis of Two-Way AF MIMO Relaying of OSTBCs with Imperfect Channel Gains\vspace*{-0.0em}}
\begin{document} 
\maketitle
\begin{abstract}
In this paper, we consider the relaying of orthogonal space time block codes (OSTBCs) in a two-way amplify-and-forward (AF) multiple-input multiple-output (MIMO) relay system with \emph{estimated} channel state information (CSI). 
A simple four phase protocol is used for training and OSTBC data transmission.
Decoding of OSTBC data at a user terminal is performed
by replacing the exact CSI by the 
estimated CSI, in a maximum likelihood decoder. 
Tight approximations for the moment generating
function (m.g.f.) of the received signal-to-noise ratio at
a user is derived under Rayleigh fading by
ignoring the higher order noise terms.
Analytical average error performance of the considered cooperative scheme is derived by using the m.g.f. expression. Moreover, the analytical diversity order of the considered scheme is also obtained for certain system configurations. It is shown by simulations and analysis that the channel estimation does not affect the diversity order of the OSTBC based two-way AF MIMO relay system. 
\end{abstract}
\vspace*{-1.0em}
\section{Introduction}
\vspace*{-0em}
Two users can exchange their information by using a relay node, in a two-way cooperative system. 
 The spectral efficiency and diversity gain of the two-way relay system can be improved by employing multiple antennas \cite{GuGoPo:08,WyOeBj:08,ArRtRk:12,ArRtRk:13,LiHo:10}. However, a key problem in two-way relay system is that the channel state information (CSI) of all links is needed at both the users, for self-interference cancellation and decoding of the data transmitted by each other.
The problem of channel estimation in two-way
relaying is not simple because each user needs to estimate the channel in between itself and relay node as well as the channel in between the other user and relay. 
Optimal channel estimation and training design for \emph{single} antenna based two-way amplify-and-forward (AF) relay networks is discussed in~\cite{GaZhLi:09}; a rate efficient two-phase training protocol for cascaded channel estimation, required for maximum likelihood (ML) detection, is proposed in this paper.
Comparison of different channel estimation schemes like cascaded channel estimation and individual channel estimation
is performed in \cite{FaSh:11}, for a two-way multiple-input multiple-output (MIMO) relay systems. The mean square error (MSE) and bit error rate (BER) performance of these two schemes are compared, in this work. 
In \cite{AlRo:10}, lower bounds on the sum rate of information transmission in
AF MIMO relay system with estimated CSI, using a precoding matrix at the relay, are obtained. 
The problem of joint source and relay optimization for
AF MIMO two-way relay systems, with imperfect CSI, is considered in \cite{JuHa:12}; 
however, no analytical performance of the AF MIMO relay system is provided.\vspace*{-2em} 
\vspace*{0.75em}
 \subsection*{Motivation and Our Contributions} 
In the existing works~\cite{AlRo:10,JuHa:12}, the transmission of the spatial multiplexed data vector, 
by using imperfect CSI based precoding matrix at the users and/or relay, is considered. These techniques require complicated
MIMO precoder designs, at different nodes; moreover, the spatial multiplexing results into
high complexity ML decoding. On the other hand, the orthogonal space time block code (OSTBC)
achieves full diversity without any channel information at the transmitter, and provides low complexity symbol wise ML decoding. 
The OSTBC based relay system does not require global CSI at the source and/or relay contrary to the existing schemes~\cite{AlRo:10,JuHa:12}; hence, it can be easily implemented in practice.
In \cite{RaRoIa:12,RaRo:11}, transmission of OSTBC in decode-and-forward (DF) protocol based two-way MIMO relay system is studied with \emph{perfect} information of the CSI, required for decoding the transmitted data at the users and relay. The schemes of \cite{RaRoIa:12,RaRo:11} avoid the need for the CSI at the transmitters in two-way MIMO relay system, contrary to~\cite{AlRo:10,JuHa:12}.  
However, due to the erroneous relaying in DF protocol, the OSTBC based two-way DF MIMO relay systems \cite{RaRoIa:12,RaRo:11} loose diversity. 
Since the AF protocol does not have this problem, it can be used for improving the OSTBC relaying in two-way MIMO relay system.
Further, the existing works \cite{RaRoIa:12,RaRo:11} do not facilitate
deriving important performance metrics such as the symbol error rate (SER) and achievable diversity, for the OSTBC based two-way MIMO relay system. 
In this paper, we address the problem of fixed gain relaying of OSTBC in a two-way AF MIMO relay system, with \emph{imperfect} CSI.
It is deduced by using the ML decoder metric of the OSTBC data that OSTBC relaying in two-way AF MIMO relaying system requires a complicated channel estimation protocol, as compared to existing two-way AF relay systems.

Our contributions in this paper are as follows: 1) Closed-form approximate expression (by ignoring higher order noise terms) for the moment generating function (m.g.f.) of the received signal-to-noise ratio (SNR) at a user in OSTBC based two-way AF MIMO relaying system with estimated CSI, is derived. 
2) The performance of the considered two-way AF MIMO cooperative system is quantified by deriving the analytical average error performance for $M$-PSK and $M$-QAM constellations. 
3) The analytical expression of diversity order of the considered system with imperfect CSI, for some special cases, is also obtained. Some useful insights into practical two-way AF MIMO 
relay system implementation are obtained by using the analytical
diversity order. 
\vspace*{-0.15in}
\section{System Model}
\begin{figure}[t!]
\centerline{\psfig{file=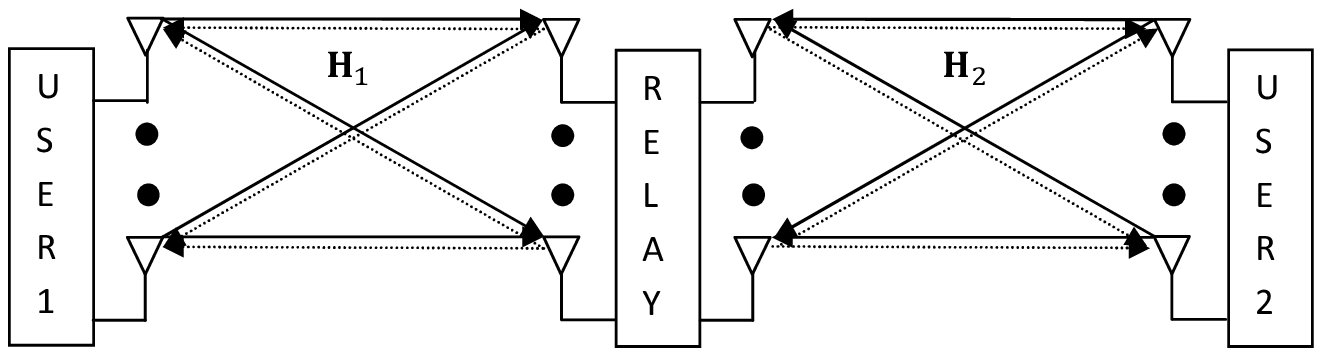,width=2.8in,height=1.0in}}\vspace*{-1mm}
\caption{Two-way MIMO relay system.}\vspace*{-6.5mm}
\label{f_1}
\end{figure}

\vspace*{-1mm}
We consider a two-way cooperative system with two users and
a single relay, as shown in Fig. \ref{f_1}, where User-$i$,
$i=1,2$, and the relay
have ${N}_{i}$ and
${N}_{r}$ antennas, respectively. Both users act as source and destination as well. 
Since the transmission of data occurs in both directions, i.e.,
to/from the relaying node, we assume that channel
reciprocity holds; so that
the reverse channel matrices are the transpose of the forward channel matrices.
If channel reciprocity does not hold, then the derived results are not applicable.
Let  $\undb{H}_{i}$
     be the $N_i \times N_r$ relay-to-User-$i$ MIMO
channel matrix containing independent and identically distributed (i.i.d.) complex circular
Gaussian elements with zero mean
and unit variance, which corresponds to the case of i.i.d. Rayleigh
fading. Transmission of data occurs in two phases;
during first phase, both users transmit their data to the relay.
In the second phase, the MIMO relay amplifies the received signals with a fixed gain, and then broadcasts to both users. There is no direct link in between the users. Throughout the paper, we assume that User-$i$ needs to decode the OSTBC data transmitted by User-$j$, $j=1,2$, $i\neq j$. 

It is assumed for analytical simplicity, that the source (User-$j$) employs
\emph{linear} OSTBC encoding. In particular, groups of $M_j$
complex symbols $c^{(j)}_1, c^{(j)}_2, . . . , c^{(j)}_{M_j}$ are mapped to an orthogonal
matrix $\undb{C}_j\in {\mathbb{C}}^{{N_j}\times T}$, $N_j \leq T$, where $T$ denotes the number
of symbol intervals used to send each OSTBC codeword. 
A detailed structure of the linear OSTBC is given in \cite[Chapter 7]{LaSt:03}. 
Since it takes $T$ symbol periods to transmit $M_j$
symbols, the code rate is $R_j = M_j/T$. 
We can write $\undb{C}_j=\left[\undb{c}_j(1), \undb{c}_j(2), \undb{c}_j(3),\dots ,\undb{c}_j(T)\right]$, where $\undb{c}_j(m)\in {\mathbb{C}}^{N_j\times 1}$, $m=1,2,...,T$,
$E\left\{\left\|\undb{c}_j(m)\right\|^2\right\}=\bar{\gamma}_j$, $E\left\{\cdot\right\}$ denotes
the expectation, $\left\|\cdot\right\|$ stands for the Euclidean norm, and $\bar{\gamma}_j$ is 
defined as the average SNR of User-$j$. 
The received signal at the relay
in the first phase is given by
   \vspace*{-0.1in}
   \beq
     \undb{Y}_{r}=\sum^{2}_{i=1}\undb{H}^{T}_{i}\undb{C}_{i}+\undb{W}_{r}, \label{che3}\vspace*{-0.1in}
      \eeq
      where $\undb{Y}_{r}\in{\mathbb{C}}^{N_r\times T}$ denotes the received signal matrix, $\left(\cdot\right)^{{T}}$ denotes the transpose; and $\undb{W}_{r}\in{\mathbb{C}}^{N_r\times T}$ with $\mbox{vec}\left(\undb{W}_{r}\right)\sim{\mathcal{CN}}\left(\undb{0}_{N_rT},\undb{I}_{N_rT}\right)$--where $\mbox{vec}\left(\cdot\right)$ is the column wise vectorization operator, $\undb{0}_{N_rT}$ represents an all zero $N_rT\times 1$ vector, and ${\mathcal{CN}}\left(\cdot,\cdot\right)$ denotes the complex 
Normal distribution--is the additive white Gaussian noise (AWGN) matrix at the relay, containing zero mean complex Gaussian
noise elements with unit variance. The relay multiplies this received signal by a constant gain matrix $\undb{G}=a\undb{I}_{N_r}$ prior to broadcasting it
to both users. In particular, the selection of $a$ ensures that an average power constraint $E\left\{\left\|\undb{G}\undb{Y}_r\right\|^2_F\right\}\leq b$,
where $\left\|\cdot\right\|_F$ is the Frobenius norm of the matrix, is satisfied at the relay. The value of $a$ is given as
\beq
a=\sqrt{\frac{b}{N_rT(1+\rho)}}, \quad b\in {\mathbb{R}}^{+}\label{a},
\eeq
where $\rho=\bar{\gamma}_i+\bar{\gamma}_j$. 
The data received at the User-$i$, during broadcasting from the relay, is given by
\vspace*{-0.1in} 
      \beq
     \undb{Y}_{i}=a\undb{H}_{i} \undb{Y}_{r}+
    {\undb{W}}_{i},\label{dir4}\vspace*{-0.1in}
    \eeq
         where ${\undb{W}}_{i}$ with $\mbox{vec}\left(\undb{W}_{i}\right)\sim{\mathcal{CN}}\left(\undb{0}_{N_iT},\undb{I}_{N_iT}\right)$ is
     the AWGN noise matrix. 
From (\ref{che3}) and (\ref{dir4}), we have 
         \beq
        \undb{Y}_i=a\undb{G}_i\undb{C}_i+a\undb{G}_{i,j}\undb{C}_j
        +a{\undb{H}}_i\undb{W}_r+{\undb{W}}_i, \label{pa00}
        \eeq
        where $\undb{G}_i=\undb{H}_i\undb{H}_i^T\in{\mathbb{C}}^{N_i\times N_i}$ and $\undb{G}_{i,j}=\undb{H}_i\undb{H}_j^T\in{\mathbb{C}}^{N_i\times N_j}$. An ML decoder of $\undb{C}_j$ can be obtained by maximizing the conditional probability density function (p.d.f.), i.e., ${f(\tilde{\undb{y}}_{i}|{\undb{g}_i},
    {\undb{g}_{i,j}},{\undb{H}}_{i},\undb{C}_{i},\undb{C}_{j})}$--where $\tilde{\undb{y}}_{i}= \undb {K}^{-1/2}_{i}\undb{y}_{i}\in{\mathbb{C}}^{N_iT\times 1}$, $\undb{y}_i=\mbox{vec}\left( \undb{Y}_i\right)\in{\mathbb{C}}^{N_iT\times 1}$, $\undb{g}_i=\mbox{vec}(\undb{G}_i)\in{\mathbb{C}}^{N^2_i\times 1}$,
   $\undb{g}_{i,j}=\mbox{vec}(\undb{G}_{i,j})\in{\mathbb{C}}^{N_iN_j\times 1}$, $\undb{K}_{i}=a^2(\undb{I}_T
   \otimes{{\undb{H}}_{i}}{{\undb{H}}^{H}_{i}})
   +\undb{I}_{N_iT}\in{\mathbb{C}}^{N_iT\times N_iT}$, $\otimes$ denotes the Kronecker product, and $\left(\cdot\right)^{{H}}$ represents the Hermitian--as  
            \begin{align}
    \hat{\undb{C}}_{j}\!\!=\!\!\arg\!\:\!\min_{\!\!\undb{C}_j} \!\!\!\;  \; \left\|
    \!\tilde{\undb{y}}_{i}\!\!-\!\!a
    \undb {K}^{-1/2}_{i}\!\!\left\{\!\!\left(\undb{C}^{T}_{i}
    \!\otimes\!\undb{I}_{N_i}\right)
    \undb{g}_i\!+\!\left(\undb{C}^{T}_{j}
    \!\otimes\!\undb{I}_{N_i}\right)
    \undb{g}_{i,j}\right\}\right\|^{2}\!\!,\,\label{dche9a}
     \end{align}
     where $\undb{A}^{1/2}$ denotes the matrix square root of matrix $\undb{A}$. 
From (\ref{dche9a}), it can be seen that we need to estimate $\undb{H}_i$ (in order to calculate $\undb{K}_i$); and cascaded channel matrices $\undb{G}_i$ and $\undb{G}_{i,j}$, to decode the OSTBC matrix $\undb{C}_j$ at User-$i$. 
\vspace*{-0.1in}
     \section{Decoding of OSTBC in Two-Way AF MIMO Relay System with Estimated CSI}
Since the proposed ML decoder (\ref{dche9a}) requires individual and cascaded channel estimates, we use a modified version of the protocol of \cite{GaZhLi:09} (which only allows for cascaded channel estimation) for channel estimation and OSTBC data transmission; 
this protocol consists of four phases. In the first phase, the relay broadcasts training data to both users. In the second phase, one of the users (User-$i$) transmits pilot symbols to the relay and relay broadcasts the received signals with unity gain. Another user (User-$j$) transmits training data and the relay broadcasts the received signals, in the third phase. Whereas, in the fourth and final phase, two-way relaying of the OSTBCs of the users takes place. 
It is assumed that the users and relay transmit unitary pilot matrices. 

In the first phase, the ML estimate of $\undb{H}_i$ at User-$i$
can
be obtained as
\vspace*{-0.1in} 
\beq
\hat {\undb{H}}_{i} = \undb{R}_{i}
{\undb{S}}_{p}^{H}\left(
{\undb{S}}_{p} {\undb{S}}^{H}_{p}\right)^{-1}
=\undb{H}_{i}+\hat{\undb{N}}_{i}\in{\mathbb{C}}^{N_i\times N_r},
\label{chanestfirstphase}\vspace*{-0.1in}
\eeq
where  $\undb{R}_{i}=\undb{H}_{i}\undb{S}_{p}+\undb{N}_{i}\in{\mathbb{C}}^{N_i\times P}$, $P=M_pN_r$, $M_p\in {\mathbb{Z}}^+$,
denotes the data matrix received
by User-$i$ due to 
the pilot matrix 
${\undb{S}}_{p}\in{\mathbb{C}}^{{N}_{r}\times P}$ broadcasted by the relay, and $\undb{N}_{i}$ with $\mbox{vec}\left(\undb{N}_{i}\right)\sim{\mathcal{CN}}\left(\undb{0}_{N_iP},\undb{I}_{N_iP}\right)$ is the AWGN noise matrix received at User-$i$. Note that
the pilot matrix ${\undb{S}}_{p}$ contains $M_p$ transmissions of an $N_r\times N_r$
unitary matrix. 
Furthermore, $\hat{\undb{N}}_{i}= \undb{N}_{i}\undb{S}^H_{p}\left(
{\undb{S}}_{p} {\undb{S}}^{H}_{p}\right)^{-1}$ with $\mbox{vec}\left(\hat{\undb{N}}_{i}\right)\sim{\mathcal{CN}}\left(\undb{0}_{N_iN_r},\undb{I}_{N_iN_r}\right)$
is the
channel estimation noise of $\undb{H}_{i}$. 

In the second phase of training, the data received at User-$j$ is given by
\vspace*{-0.1in}
 \beq
\tilde{\undb{R}}_{j}=\undb{H}_{j}\undb{H}^T_{i}\undb{C}_{p_{i}}+ \undb{H}_{j}\undb{U}_i+\undb{N}_j,\vspace*{-0.1in}
  \label{j22}
    \eeq
     where ${\undb{C}}_{p_i}
\in{\mathbb{C}}^{N_{i}\times L}$, $L=N_{p_i}N_i$, $N_{p_i}\in {\mathbb{Z}}^+$, denotes the unitary
training matrix transmitted by User-$i$; 
${\undb{U}}_{i}$ with $\mbox{vec}\left({\undb{U}}_{i}\right)\sim{\mathcal{CN}}\left(\undb{0}_{N_rL},\undb{I}_{N_rL}\right)$ and ${\undb{N}}_{j}$ with $\mbox{vec}\left({\undb{N}}_{j}\right)\sim{\mathcal{CN}}\left(\undb{0}_{N_jL},\undb{I}_{N_jL}\right)$ represent AWGN noise received at the relay and User-$j$, respectively. By applying the vectorization operator over (\ref{j22}) and performing whitening of the noise, we obtain $\breve{\undb{r}}_{j}={\undb{K}}_j^{-1/2}\tilde{\undb{r}}_{j}=
{\undb{K}}_j^{-1/2}\left(\undb{C}^{T}_{p_i}
  \!\otimes\!\undb{I}_{N_j}\right)\undb{g}_{j,i}+ \undb{n}_{j,i}$, where $\tilde{\undb{r}}_{j}=\mbox{vec}\left(\tilde{\undb{R}}_{j}\right)$, ${\undb{K}}_j=\left(\undb{I}_L
\otimes{\undb{H}}_{j}{\undb{H}}^{H}_{j}\right)+
\undb{I}_{LN_j}$, and $\undb{n}_{j,i}\sim{\cal CN}\left(\undb{0}_{N_jL}, 
    \undb{I}_{N_jL}\right)$ is the AWGN noise vector. By minimizing $\left\|\breve{\undb{r}}_{j}-
{\undb{K}}_j^{-1/2}\left(\undb{C}^{T}_{p_i}
  \!\otimes\!\undb{I}_{N_j}\right)\undb{g}_{j,i}\right\|^2$ with respect to (w.r.t.) ${\undb{g}}_{j,i}=\mbox{vec}\left\{{\undb{G}}_{j,i}\right\}=\mbox{vec}\left\{\undb{H}_{j}\undb{H}^T_{i}\right\}\in{\mathbb{C}}^{N_iN_j\times 1}$, we get the estimate of ${\undb{g}}_{j,i}$ as
\begin{align}
  \!\hat{\undb{g}}_{j,i}\!\!=\!\!\left(\!
(\undb{C}^{*}_{p_i}\otimes\undb{I}_{N_j})\tilde{\undb{K}}_j^{-1}(\undb{C}^{T}_{p_i}
  \!\otimes\!\undb{I}_{N_j})\!\right)^{-1}\!\!\!\!\!\!\!\! (\undb{C}^{*}_{p_i}
  \!\otimes\!\undb{I}_{N_j})\tilde{\undb{K}}_j^{-1}
   \mbox{vec}(\tilde{\undb{R}}_{j})
 ,\vspace*{-0.1in}\label{j44}
    \end{align}
where $\tilde{\undb{K}}_j=\left(\undb{I}_L
\otimes\hat{\undb{H}}_{j}\hat{\undb{H}}^{H}_{j}\right)+
\undb{I}_{LN_j}$ and $\left(\cdot\right)^{{*}}$ denotes the matrix conjugate. 
By substituting the value of $\tilde{\undb{r}}_{j}$ from (\ref{j22}) to (\ref{j44}),  and after some algebraic manipulations, we get
    \beq
    \hat{\undb{g}}_{j,i}=\undb{g}_{j,i}+\left(\undb{C}^{*}_{p_i}
    \otimes{\undb{I}_{N_j}}\right)\tilde{\undb{K}}^{1/2}_{j}
     {\undb{n}}_{j,i}.\label{che2g} \vspace*{-0.05in} 
    \eeq
The data received at User-$i$,
in the second phase, is given by\vspace*{-0.05in}
     \beq
    \bar{\undb{R}}_{i}= \undb{G}_{i}\undb{C}_{p_i}+
    \undb{H}_{i} \undb{U}_{j}+{\undb{N}}_{i},\, \label{che2a}\vspace*{-0.05in}
    \eeq
    where ${\undb{U}}_{j}$ with $\mbox{vec}\left({\undb{U}}_{j}\right)\sim{\mathcal{CN}}\left(\undb{0}_{N_rL},\undb{I}_{N_rL}\right)$ and ${\undb{N}}_{i}$ with $\mbox{vec}\left({\undb{N}}_{i}\right)\sim{\mathcal{CN}}\left(\undb{0}_{N_iL},\undb{I}_{N_iL}\right)$ represent AWGN noise received at the relay and User-$i$, respectively. 
    From (\ref{che2a}), after some manipulations, the ML estimate of $\undb{g}_{i}$ 
    can be obtained as\vspace*{-0.025in}
    \beq
    \hat{\undb{g}}_i=\undb{g}_{i}+\left(\undb{C}^{*}_{p_i}
    \otimes{\undb{I}_{N_i}}\right)\tilde{\undb{K}}^{1/2}_{i}
     {\undb{n}}_{i,i},\label{che2f}\vspace*{-0.05in}  
    \eeq
      where 
    $\undb{n}_{i,i}\sim{\cal CN}\left(\undb{0}_{N_iL},
    \undb{I}_{N_iL}\right)$ is the AWGN noise vector. Similarly, 
 the channel estimates, $\hat{\undb{g}}_{i,j}$ and $\hat{\undb{g}}_{j}$, in the third phase of training, can be
 obtained.     
By replacing the
exact channel gains with their estimated values in (\ref{dche9a}), we
get a decoder of the OSTBC matrix $\undb{C}_{j}$ at User-$i$:
\vspace*{-0.00in}
 \begin{align}
    \hat{\undb{C}}_{j}\!\! =\!\arg\:\!\!\min_{\undb{C}_j}\!\!\! \; & \; \left\|
    \hat{\tilde{\undb{y}}}_{i}\!\!-\!\!a
   \hat{\undb {K}}^{-1/2}_{i}\!\!\left\{\!\!\left(\undb{C}^{T}_{i}
    \!\otimes\!\undb{I}_{N_i}\!\right)
    \hat{\undb{g}}_i\!\!+\!\!\left(\undb{C}^{T}_{j}
    \!\otimes\!\undb{I}_{N_i}\right)
 \hat{\undb{g}}_{i,j}\right\}\right\|^{2},\,\vspace*{-0.1in}\label{de1}
     \end{align}\vspace*{-0.0in}
     \hspace*{-0.3em}where $\hat{\tilde{\undb{y}}}_{i}=\hat{\undb{K}}_{i}^{-1/2}{\undb{y}}_{i}$,
     $\hat{\undb{K}}_{i}=a^2(\undb{I}_T
   \otimes{\hat{\undb{H}}_{i}}{\hat{\undb{H}}^{H}_{i}})
   +\undb{I}_{N_iT}$, $\hat{\undb{g}}_i$ is given in (\ref{che2f}), and
   $\hat{\undb{g}}_{i,j}$ can be obtained from (\ref{che2g}) by swapping $i$ and $j$.
From~(\ref{de1}) and~\cite[Eq.~(7.4.2)]{LaSt:03}, and after some algebra, we get the following symbol-wise decoder of the OSTBC data, in channel estimation based two-way AF MIMO relay system:
    \beq
\hat{c}^{(j)}_n\!=\!\min_{c^{(j)}_n} \left|c^{(j)}_n\!\!-\!\!\frac{\mbox{Re}\left(\!\mbox{Tr}\left\{\breve{\undb{Y}}^H_i\undb{X}\undb{A}_n\right\}\!\right)\!\!-\!\!\jmath \mbox{Im}\left(\!\mbox{Tr}\left\{\breve{\undb{Y}}^H_i\undb{X}\undb{B}_n\right\}\!\right)}{\left\|\undb{X}\right\|^2}\right|^2\!\!\!\!.\label{n2}
    \eeq
 In (\ref{n2}), $n=1,2,..,M_j$, $\undb{A}_n$ and $\undb{B}_n$ are $N_j\times T$ OSTBC specific dispersion matrices given in~\cite[Eq. (7.1.1)]{LaSt:03}, $\mbox{Re}\left(\cdot\right)$ and $\mbox{Im}\left(\cdot\right)$ represent real and imaginary parts, respectively, of a complex quantity, $\mbox{Tr}\left\{\cdot\right\}$ denotes the matrix trace, $\jmath=\sqrt{-1}$,  $\undb{X}=\left(\hat{\undb{H}}_i\hat{\undb{H}}^H_i+\undb{I}_{N_i}\right)^{-1/2}\hat{\undb{G}}_{i,j}$, and
    $\mbox{vec}\left(\breve{\undb{Y}}_i\right)=\hat{\tilde{\undb{y}}}_i-\left(\undb{C}^T_i\otimes 
     \left(\hat{\undb{H}}_i\hat{\undb{H}}^H_i+\undb{I}_{N_i}\right)^{-1/2}\right)\hat{\undb{g}}_{i}$.\vspace{-.1in}
\section{Performance Analysis}\vspace{-.01in}
        We analyze the performance of
        the two-way AF MIMO relay system with estimated
        channel gains,
        in this section.  
It is assumed that
$\undb{C}_{p_i}\undb{C}^H_{p_i}=\undb{I}_{N_i}$ and 
$\undb{C}_i\undb{C}^H_i=\undb{I}_{N_i}$.
Let $\undb{b}\left(k\right)$ denotes the $k$-th column of
a matrix $\undb{B}$; therefore, we can write the $m$-th column of $\undb{Y}_i$
from (\ref{pa00}) as
\vspace*{-0.06in}
\beq
\undb{y}_i(m)\!=\!a\undb{G}_i\undb{c}_i(m)\!+\!a\undb{G}_{i,j}\undb{c}_j(m)
        \!+\!a{\undb{H}}_i\undb{w}_r(m)\!+\!{\undb{w}}_i(m). \label{pa01}\vspace*{-0.06in}
\eeq
From (\ref{pa01}), it can be seen that the overall 
input-output relationship is equivalent to 
multiple access channels based two-user
MIMO system with effective channel matrices $\undb{G}_i$ and 
$\undb{G}_{i,j}$; and 
additive colored Gaussian noise, $a{\undb{H}}_i\undb{w}_r(m)+{\undb{w}}_i(m)$, which has
the following conditional covariance:
\vspace*{-0.1in}
\beq
\!\!\!\!\!\!\breve{\undb{K}}_i\!=\!{E}\!\left\{\!\left(a{\undb{H}}_i\undb{w}_r(m)\!+\!{\undb{w}}_i(m)\!\right)\!\!\left(a{\undb{H}}_i\undb{w}_r(m)\!+\!{\undb{w}}_i(m)\right)^H\!\!|\undb{H}_i\!\right\}. \label{pa021}\vspace*{-0.0in}
\eeq
Therefore,
the ML decision variable at User-$i$ will be
\vspace*{-0.1in}
\beq
 \Lambda\!=\!\min_{\tilde{\undb{c}}_j} \left\|\tilde{\undb{y}}_i(m)\!-\!a\breve{\undb{K}}^{-1/2}_i\left(\undb{G}_i\undb{c}_i(m)\!+\!\undb{G}_{i,j}\tilde{\undb{c}}_j(m)\right)\right\|^2,\label{pa02}\vspace*{-0.1in}
 \eeq
where $\tilde{\undb{y}}_i(m)=\breve{\undb{K}}^{-1/2}_i{\undb{y}}_i(m)$. 
     It can be observed from (\ref{pa02})
     that the decision variable depends upon the
     exact value of the effective channel matrices.
    However, the
    destination user utilizes estimated channel matrices 
    (gained during the training phases),
    in place of the exact channel matrices
    in the decision variable. Therefore,
    from (\ref{pa02}), the decision variable with estimated channel
    matrices can be written as 
    \beq
 \Lambda\!=\!\min_{\tilde{\undb{c}}_j} \left\|\hat{\tilde{\undb{y}}}_i(m)\!-\!a\hat{\breve{\undb{K}}}^{-1/2}_i\!\left(\!\hat{\undb{G}}_i\undb{c}_i(m)
 \!+\!\hat{\undb{G}}_{i,j}\tilde{\undb{c}}_j(m)\right)\right\|^2,\label{pa022}
 \eeq
 where $\hat{\undb{G}}_i$ and $\hat{\undb{G}}_{i,j}$ denote the estimate of ${\undb{G}}_i$ and ${\undb{G}}_{i,j}$, respectively; $\hat{\tilde{\undb{y}}}_i(m)=\hat{\breve{\undb{K}}}^{-1/2}_i{\undb{y}}_i(m)$, and
\beq
\!\!\!\!\!\hat{\breve{\undb{K}}}_i\!\!=\!{E}\!\!\left\{\!\!\left(\!a\hat{\undb{H}}_i\undb{w}_r(m)\!+\!{\undb{w}}_i(m)\!\!\right)\!\!\left(a\hat{\undb{H}}_i\undb{w}_r(m)\!+\!{\undb{w}}_i(m)\right)^H\!\!\!\!|\undb{H}_i\!\!\right\}\label{pa023}
\eeq 
is the estimate of ${\breve{\undb{K}}}_i$.
From~(\ref{chanestfirstphase}) and (\ref{pa023}), we have
\begin{align}
\label{pa024}
\hat{\breve{\undb{K}}}_i={E}\left\{\left(a{\undb{H}}_i\undb{w}_r(m)+a\hat{\undb{N}}_i\undb{w}_r(m)+{\undb{w}}_i(m)\right)\right.\nonumber\\
\times\left.\left(a{\undb{H}}_i\undb{w}_r(m)+a\hat{\undb{N}}_i\undb{w}_r(m)+{\undb{w}}_i(m)\right)^H|\undb{H}_i\right\}.
\end{align}
By ignoring the higher order noise in (\ref{pa024}), we have $\hat{\breve{\undb{K}}}_i\cong{\breve{\undb{K}}}_i$; hence, $\hat{\tilde{\undb{y}}}_i(m)\cong{\breve{\undb{K}}}^{-1/2}_i{\undb{y}}_i(m)$. Since the higher order noise terms have very small value as compared to other noise terms, these terms can be ignored for simplifying the analysis. It is shown in \cite{Manav:07, Manav:08, Manav:13}
that this approximation works very
well for performance analysis of different communication systems. Moreover, it will be verified by simulation in Section V that the analytical results, derived under this assumption, are very accurate. 

Let us observe that $\hat{\undb{g}}_i=\mbox{vec}\left(\hat{\undb{G}}_i\right)$ and $\hat{\undb{g}}_{i,j}=\mbox{vec}\left(\hat{\undb{G}}_{i,j}\right)$; therefore,
we can write $\hat{\undb{G}}_{i,j}$
 and $\hat{\undb{G}}_i$, in terms of ${\undb{G}}_{i,j}$ and ${\undb{G}}_i$, by using (\ref{che2g}) and (\ref{che2f}), respectively, after some straightforward algebra. Substituting these values of $\hat{\undb{G}}_i$ and $\hat{\undb{G}}_{i,j}$, and 
value of ${\undb{y}}_{i}(m)$ (given in (\ref{pa01})) in (\ref{pa022}); 
    and after some algebra, the decision variable at the User-$i$ can be written as
    \beq
 \Lambda=\min_{\tilde{\undb{c}}_j} \left\|a{\breve{\undb{K}}}^{-1/2}_i\undb{G}_{i,j}\left(\undb{c}_j(m)-\tilde{\undb{c}}_j(m)\right)+\undb{w}(m)\right\|^2,\label{pa03}\vspace*{-0.5em}
 \eeq
 where $\undb{w}(m)=\breve{\undb{K}}^{-1/2}_i\left\{{a\undb{H}}_i\undb{w}_r(m)+{\undb{w}}_i(m)-a\Big(\hat{\undb{n}}_{i,i}(m)\right.$ $\left.\undb{c}_i(m)
 +\hat{\undb{n}}_{i,j}(m)\undb{c}_j(m)
 \Big)\right\}$; $\hat{\undb{n}}_{i,i}(m)$ and $\hat{\undb{n}}_{i,j}(m)$
 are the estimation noises of $\undb{g}_{i}(m)$ and $\undb{g}_{i,j}(m)$,
 respectively. From (\ref{pa03}), we can write the ML decision variable for 
 transmitted codeword $\undb{C}_j$ at User-$i$ as
 \beq
 \Lambda=\min_{\tilde{\undb{C}}_j} \left\|a\breve{\undb{K}}^{-1/2}_i\undb{G}_{i,j}\left(\undb{C}_j-\tilde{\undb{C}}_j\right)+\undb{W}\right\|^2.\label{pa6}
 \eeq
 In (\ref{pa6}), $\undb{W}=\breve{\undb{K}}^{-1/2}_i\!\!\left\{{a\undb{H}}_i\undb{W}_r\!+\!{\undb{W}}_i\!-\!a\left(\hat{\undb{N}}_{i,i}\undb{C}_i\!+\!\hat{\undb{N}}_{i,j}\undb{C}_j
 \right)\right\}$ is the additive Gaussian noise ($\hat{\undb{N}}_{i,i}\in{\mathbb{C}}^{N_i\times N_i}$ and
       $\hat{{\undb{N}}}_{i,j}\in{\mathbb{C}}^{N_i\times N_j}$
 being the estimation noises of $\undb{G}_{i}$ and $\undb{G}_{i,j}$,
 respectively) present in the decision variable with conditional covariance matrix\vspace*{2 mm}
  \begin{align}
  \label{pa9}
 &\check{\undb{K}}_{i}=\!{E}\Big\{\!\breve{\undb{K}}^{-1/2}_i
\left(a\undb{H}_i\undb{W}_r\!+\!{\undb{W}}_i\!-\!a\left(\hat{\undb{N}}_{i,i}\undb{C}_i\!+\!\hat{\undb{N}}_{i,j}\undb{C}_j\right)\right)\nonumber\\
&\times\left(a\undb{H}_i\undb{W}_r\!+\!{\undb{W}}_i\!-\!a\left(\hat{\undb{N}}_{i,i}\undb{C}_i\!+\!\hat{\undb{N}}_{i,j}\undb{C}_j\right)\right)^{H}\breve{\undb{K}}^{-H/2}_i|{\undb{H}}_i\Big\}\nonumber\\
&=\breve{\undb{K}}^{-1/2}_i\left(Z_1\undb{I}_{N_i}+Z_2\undb{H}_i\undb{H}^{H}_i\right)\breve{\undb{K}}^{-H/2}_i,
\end{align}
where $Z_1=1+a^2\left(1/N_{p_i}+1/N_{p_j}\right)$ and $Z_2=a^2(1+1/N_{p_i}+1/N_{p_j})$. 
  From
 (\ref{pa6}), the instantaneous SNR of the
       $n$-th 
       symbol $c^{(j)}_{n}$
       encoded in the OSTBC $\undb{C}_{j}$, can be
       written after linear OSTBC processing, using~\cite[Eq.~(11)]{ChWiZa:03}, 
       and after some algebraic manipulations, as
        \begin{align}
    \gamma^{(i)}_{n}&=\left\|a
    \check{\undb{K}}^{-1/2}_{i}\breve{\undb{K}}^{-1/2}_{i}
                      {\undb{G}_{i,j}}
    \right\|^{2}_{F}{E}\left\{\left|c^{(j)}_{n}\right|^{2}\right\}\nonumber\\
    &=
  {\alpha_j\bar{\gamma}_ja^2}\mbox{Tr}\left\{{\undb{G}^{H}_{i,j}} \left(Z_1\undb{I}_{N_i}+Z_2\undb{H}_i\undb{H}^{H}_i\right)^{-1}{\undb{G}_{i,j}}\right\}
   ,\label{pa10}
     \end{align}
     here 
     $\alpha_j=1/(R_jN_{j})$. 
  {Since the right hand side of (\ref{pa10})
     is independent of $n$, hence, we drop subscript
     $n$; and 
     denote the received instantaneous SNR at User-$i$ as $\gamma^{(i)}$,
      in rest of the paper. 
%
\vspace*{-0.1in}
      \subsection{Expression of M.G.F.}
    The following theorem provides a closed-form expression
    for the m.g.f. of the SNR $\gamma^{(i)}$. 
\begin{theorem}
\label{the:mgf1}
      The m.g.f. of $\gamma^{(i)}$ is given as\vspace*{0.5em}
                   \begin{align}
             \label{pa11}
      M_{\gamma^{(i)}}(s)=\kappa^{-1}
      \det\left({\undb{J}(s)}\right)
      ,
     \end{align}
      where $\det\left(\cdot\right)$ denotes the determinant, $\kappa=\prod^{q}_{l=1}\Gamma{(p-l+1)}
      \Gamma{(q-l+1)}$, $p=\max(N_{r},N_{i}),
      q=\min(N_{r},N_{i})$,
  and $\undb{J}(s)$ is
    the $q \times q$ Hankel matrix with $(t,v)$-th entry
      \begin{align}
{J_{tv}}(s)&=\!\frac{\Gamma(\nu_{tv})}{(\frac{Z_2}{Z_1})^{\nu_{tv}}(1\!+\!\frac{a^2s\alpha_j\bar{\gamma}_j}{Z_2})^{\nu_{tv}+N_j}}\sum^{N_{j}}_{k=0}
\binom{N_j}{k}
\left(\frac{a^2s\alpha_j\bar{\gamma}_j}{Z_2}\right)^{k}\nonumber\\
&\times U\left(\nu_{tv},\nu_{tv}+1-k,\frac{Z_1}
{Z_2\left(1+\frac{a^2s\alpha_j\bar{\gamma}_j}{Z_2}\right)}\right),
         \label{pa22}
             \end{align}\\
    where $\nu_{tv}=t+v+p-q-1$, $\Gamma(\cdot)$
    is the Gamma function, and $U(\cdot,\cdot,\cdot)$ is
    the confluent hypergeometric function of the second kind.
      \end{theorem}
     Proof: A proof of Theorem 1 is given in Appendix I.

     It can be seen from (\ref{pa11}) and (\ref{pa22})
     that the m.g.f. has a direct dependence
     on $N_j$; however, the dependence on $N_r$ and $N_i$ is
     through their maximum and minimum values,
     $p$ and $q$, respectively. 
     \vspace*{-0.1in}
\subsection{Average Error Performance}
\subsubsection{$M$-PSK Constellation}
     The SER of User-$i$
for the $M$-PSK constellation is given
by~\cite{SiAl:05}\vspace*{-2mm}
\beq
     P^{(i)}_{MPSK}=\frac{1}{\pi}\int^{\theta}_{0}{M_{\gamma^{(i)}}\left(\frac{g_{MPSK}}{\sin^{2}{
     \theta}}\right)}d\theta,\label{pa24}\vspace*{-0em}
     \eeq
where $\theta=\pi(M-1)/M$ and $g_{MPSK}=
\sin^{2}({{\pi}/{M}})$. Therefore, the
SER of the channel estimation based two-way AF MIMO relay
system 
can be calculated from~(\ref{pa11}) and (\ref{pa24}).
The integral in~(\ref{pa24}) can be
numerically computed with the help of MATLAB.

From (\ref{pa24}), for BPSK ($M=2$), we can write the BER of User-$i$ as
\beq
P^{(i)}_{BPSK}=\frac{1}{\pi}\int^{\pi/2}_{0}M_{\gamma^{(i)}}\left(\frac{1}{\mbox{sin}^2\theta}\right).\label{B1}
\eeq
For $q=1$ ($q$ is defined after (\ref{pa11})), from (\ref{pa24}), (\ref{pa19}), and (\ref{pa19a}), we have
\begin{align}
\label{B2}
&P^{(i)}_{BPSK}=\frac{1}{\Gamma(p)}\int^{\infty}_{0}\lambda^{p-1}\mbox{exp}(-\lambda)\nonumber\\
&\times\left\{\frac{1}{\pi}\int^{\pi/2}_{0}\!\!
{\left(1+\frac{a^2\alpha_j\bar{\gamma}_j\lambda}{Z_1(1+\frac{Z_2}{Z_1}\lambda)\mbox{sin}^2\theta}\right)}^{-1}{d\theta}\right\}
d\lambda.
\end{align}\\
With help of \cite{SiAl:05}, the inner integral in (\ref{B2}) can be solved as
\begin{align}
&\int^{\pi/2}_{0}
{\left(1+\frac{a^2\alpha_j\bar{\gamma}_j\lambda}{Z_1(1+\frac{Z_2}{Z_1}\lambda)\mbox{sin}^2\theta}\right)}^{-1} {d\theta}\nonumber\\
=
&\frac{1}{2}\left(\!1\!-\!\sum^{N_j-1}_{k=0}\frac{(^{2k}_{k})(1+\frac{Z_2}{Z_1}\lambda)^k\sqrt(a^2\alpha_j\bar{\gamma}_j\lambda/Z_1)}{4^k(1+\frac{Z_2}{Z_1}(1+a^2\alpha_j\bar{\gamma}_j/Z_2)\lambda)^{k+1/2}}\right).\label{B3}
\end{align}\\
From (\ref{B2}) and (\ref{B3}), using binomial series expansion of the term $(1+\frac{Z_2}{Z_1}\lambda)^k$, and with help of \cite[Eq. (3.383.5)]{GrRy:03}, we get
 \begin{align}
&\!\!P^{(i)}_{BPSK}\!\!=\!\!\frac{1}{2}\!\!\left(\!\!1\!\!-\!\!\!\!\sum^{N_j-1}_{k=0}\!\!\sum^{k}_{l=0}\!\!\frac{(^{2k}_{k})(^{k}_{l})\Gamma{(p+l+1/2)}\sqrt{a^2\alpha_j\bar{\gamma}_j/Z_2}}{\left(\frac{Z_2}{Z_1}\right)^{p}(1+a^2\alpha_j\bar{\gamma}_j/Z_2)^{p+l+1/2}4^k}\right.\nonumber\\
&\left.\times\frac{1}{\Gamma(p)} U\left(p\!+\!l\!+\!1/2;p\!+\!l\!-\!k\!-\!1;\frac{Z_1}{(Z_2+a^2\alpha_j\bar{\gamma}_j)}\right)\!\right).\label{B4}
 \end{align}\\
 \subsubsection{$M$-QAM Constellation}
 The SER of User-$i$
for the $M$-QAM constellation can be obtained as~\cite{LeHaMa:12}
\begin{align} &P^{(i)}_{MQAM}=\frac{4}{\pi}\left(1-\frac{1}{\sqrt{M}}\right)\int^{\pi/2}_{0}{M_{\gamma^{(i)}}\left(\frac{g_{MQAM}}{\sin^{2}{
     \theta}}\right)}d\theta\nonumber\\
     &-\frac{4}{\pi}\left(1-\frac{1}{\sqrt{M}}\right)^2\int^{\pi/4}_{0}{M_{\gamma^{(i)}}\left(\frac{g_{MQAM}}{\sin^{2}{
     \theta}}\right)}d\theta,\label{pa24aa}\vspace*{-0em}
   \end{align}
     where $g_{MQAM}=3/\left(2(M-1)\right)$. The SER for the $M$-QAM constellation at User-$i$
     can be found by using~(\ref{pa11}) and (\ref{pa24aa}). The finite integrals in (\ref{pa24aa})
     can be evaluated numerically. 
\subsection{Diversity Order }
In the high
SNR region, the diversity order is a key parameter,
which dictates the system performance.
This parameter can be derived by
focusing on the asymptotic expansion of the m.g.f.
of the instantaneous received SNR. 
By using~(\ref{pa11}), we can
find the diversity order of the OSTBC based two-way AF
MIMO relay system with imperfect channel knowledge,
for some special cases, as given in Theorem 2.
\begin{theorem}\label{thedecdiv1}
 The diversity order 
 of the channel estimates based
 two-way AF MIMO relay system is 
  $p$, if $p<N_j$; and $N_j$, if $p>N_j$; for $\mbox{min}(N_{r},N_{i})= 1$.
\end{theorem}
Proof: A proof of Theorem 2 is given in Appendix II.
\vspace*{-2mm}
\begin{figure}[t!]
\vspace*{-5mm}
\vspace*{-3mm}
\centerline{\psfig{file=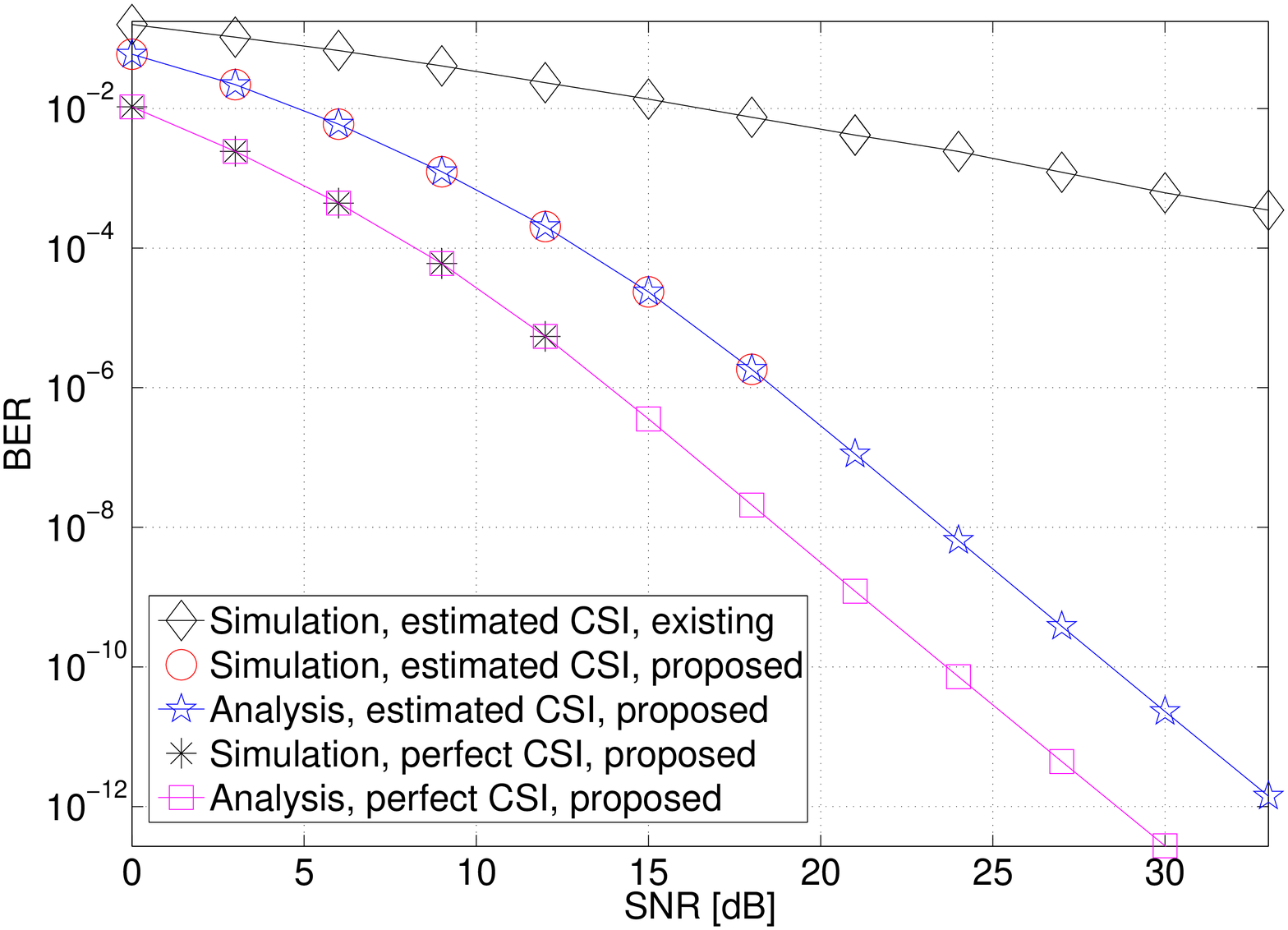,width=3.5in,height=2.5in}}\vspace*{-5mm}
\caption{Analytical and simulated BER plots of the two-way AF MIMO relay system with $N_{1}=N_{2}=N_r=2$, $N_{p_1}=N_{p_2}=1$, $a=1$, Alamouti code, perfect and estimated CSI, and BPSK constellation; and simulated BER plots of the existing two-way AF relay system with estimated CSI~\cite{GaZhLi:09}.}\vspace*{-7mm}
\label{f_5}
\end{figure}
\vspace*{0.0in}
\section{Simulation and Analytical results}
\vspace*{-.05in}
                \label{sim}
A three node two-way cooperative system shown in Fig.~\ref{f_1}
is considered for simulation and analysis. We present numerical results for
$\bar{\gamma}_1=\bar{\gamma}_2=\bar{\gamma}$, which we call the SNR,
in Figs.~\ref{f_5},~\ref{f_6},~\ref{f_10}, and \ref{f_7}.
All results are shown for
one of the two users.

In Fig.~\ref{f_5}, simulation results
   of the two-way AF MIMO relay system
   using the ML decoder with perfect CSI (given in (\ref{dche9a})) 
   and imperfect CSI (given in (\ref{de1}))
   are obtained for $N_{1}=N_{2}=N_{r}=2$, $N_{p_1}=N_{p_2}=M_p=1$, $a=1$, Alamouti
    OSTBC, and BPSK constellation.
It can be seen from Fig.~\ref{f_5} that
the ML decoder with perfect CSI outperforms
the ML decoder with imperfect CSI
at all SNRs
considered in the figure.
For example, for BER=$10^{-6}$,
the ML decoder looses approximately
5~dB due to the usage of imperfect CSI. 
Moreover, Fig.~\ref{f_5} also presents the 
analytical BER curve of the two-way
AF MIMO relaying with perfect and imperfect CSI for BPSK
modulation; analytical curves are generated based on the 
closed-form
expression, given in (\ref{B4}).
 It can be noticed from
 Fig.~\ref{f_5}, that analytical and simulated curves are closely matched
 for all SNR values considered in the simulation; this indicates
 that neglecting higher order noise does not lead to any
 serious degradation in the analytical performance. 
 Further, the slopes of the analytical BER versus
SNR plots 
at
high SNR in Fig.~\ref{f_5} indicate that
the diversity order
of the considered two-way AF MIMO relay system 
with perfect or imperfect CSI is four;
therefore, the quality of CSI does not affect 
the diversity
order of the ML decoder of two-way AF MIMO relay system
with estimated CSI. We have also plotted the simulated BER
performance of the same rate existing optimal training based 
AF two-way relaying system~\cite{GaZhLi:09},
in Fig.~\ref{f_5}. It can be seen from Fig.~\ref{f_5}
that the considered training based two-way AF MIMO relay system significantly outperforms
the existing single antenna and optimal training based two-way AF scheme~\cite{GaZhLi:09}.
Therefore, the proposed four phase training protocol (though appears to be rate deficient) enables the
OSTBC based two-way AF MIMO relay system to significantly outperform the same rate
single antenna based two-way AF relaying scheme which uses optimal training sequences.  

\begin{figure}[t!]
\vspace*{-2em}
\centerline{\psfig{file=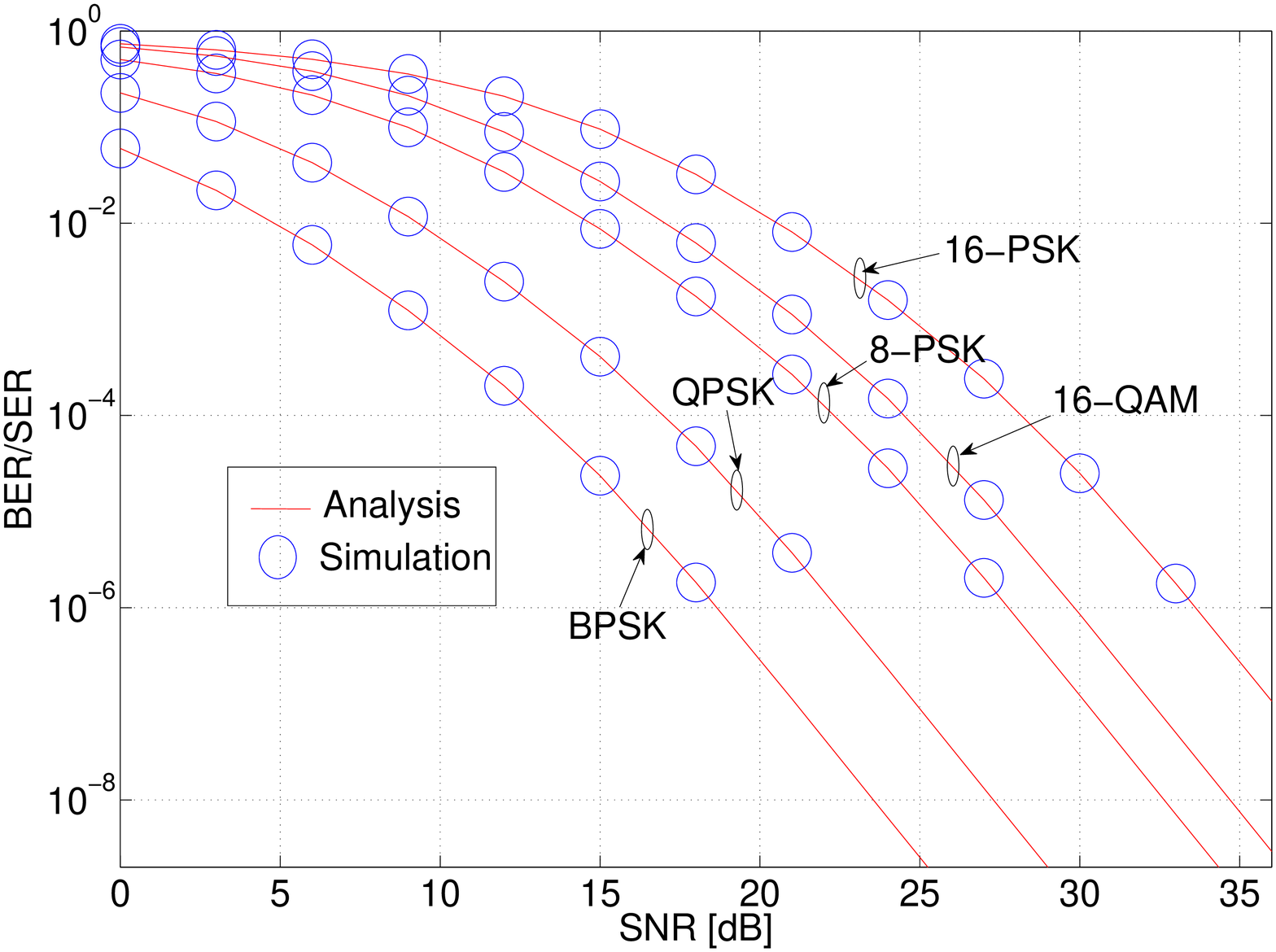,width=3.5in,height=2.5in}}\vspace*{-5mm}
\caption{Analytical and simulated BER/SER plots of the two-way AF MIMO relay system with $N_{1}=N_{2}=N_r=2$, $N_{p_1}=N_{p_2}=1$, $a=1$, Alamouti code, BPSK, QPSK, 8-PSK, 16-PSK, and 16-QAM.}\vspace*{-6mm}
\label{f_6}
\end{figure}
The analytical and simulated performance
of the ML decoder is obtained
for $N_{1}=N_{2}=N_{r}=2$, $N_{p_1}=N_{p_2}=1$, $a=1$, Alamouti STBC,
BPSK, QPSK, 8-PSK, 16-PSK, and 16-QAM,
in Fig.~\ref{f_6}. 
The simulation
results closely follow the analytical BER/SER values,
as seen in Fig.~\ref{f_6}.
In Fig. \ref{f_10}, we have plotted the simulated 
and analytical 
error performance
of the ML decoder with estimated CSI for $N_{p_1}=N_{p_2}=M_p=1$,
$N_{1}=N_{2}=2$, $N_r=1$, $a=1$, BPSK, and QPSK. 
It can be seen from Fig. \ref{f_10} that the simulated
and
analytical values of the BER/SER have a close match. 
Further, the diversity of the considered two-way AF
MIMO relay system is two as seen in Fig. \ref{f_10};
this observation corroborates the analytical diversity
order given in Theorem~2. 

We have plotted analytical BER of the two-way AF MIMO relay
system with $N_{1}=N_{2}=N_r=2$; $N_p=N_{p_1}=N_{p_2}=1,2,4,8,16$, $a=1$, perfect CSI, 
BPSK constellation, and Alamouti code. It can be seen from 
Fig. \ref{f_7} that the BER performance of the ML decoder
using estimated CSI improves with increasing number of
training blocks. For $N_p=16$, the ML decoder with estimated
CSI performs very close to the ML decoder with perfect CSI,
as seen in Fig. \ref{f_7}. 
\vspace*{-0.1in}
\section{Conclusions}\vspace*{-1mm}
This paper has discussed the transmission of OSTBC
in a two-way AF MIMO relay system with estimated channel gains.
We have investigated the statistical properties of instantaneous 
received SNR of a user, utilizing an ML decoder with imperfect CSI, by using the tools of finite-dimensional random matrix
theory.
The performance of the scheme has been discussed in terms of BER, SER, and
diversity order
for some special cases, by using the tight approximations of m.g.f.;
simulation and analytical results have demonstrated that the
diversity order of the two-way AF MIMO relay system
is not affected by the
use of
estimated channel matrices.
\begin{figure}[t!]
\vspace*{-5mm}
\vspace*{-3mm}
\centerline{\psfig{file=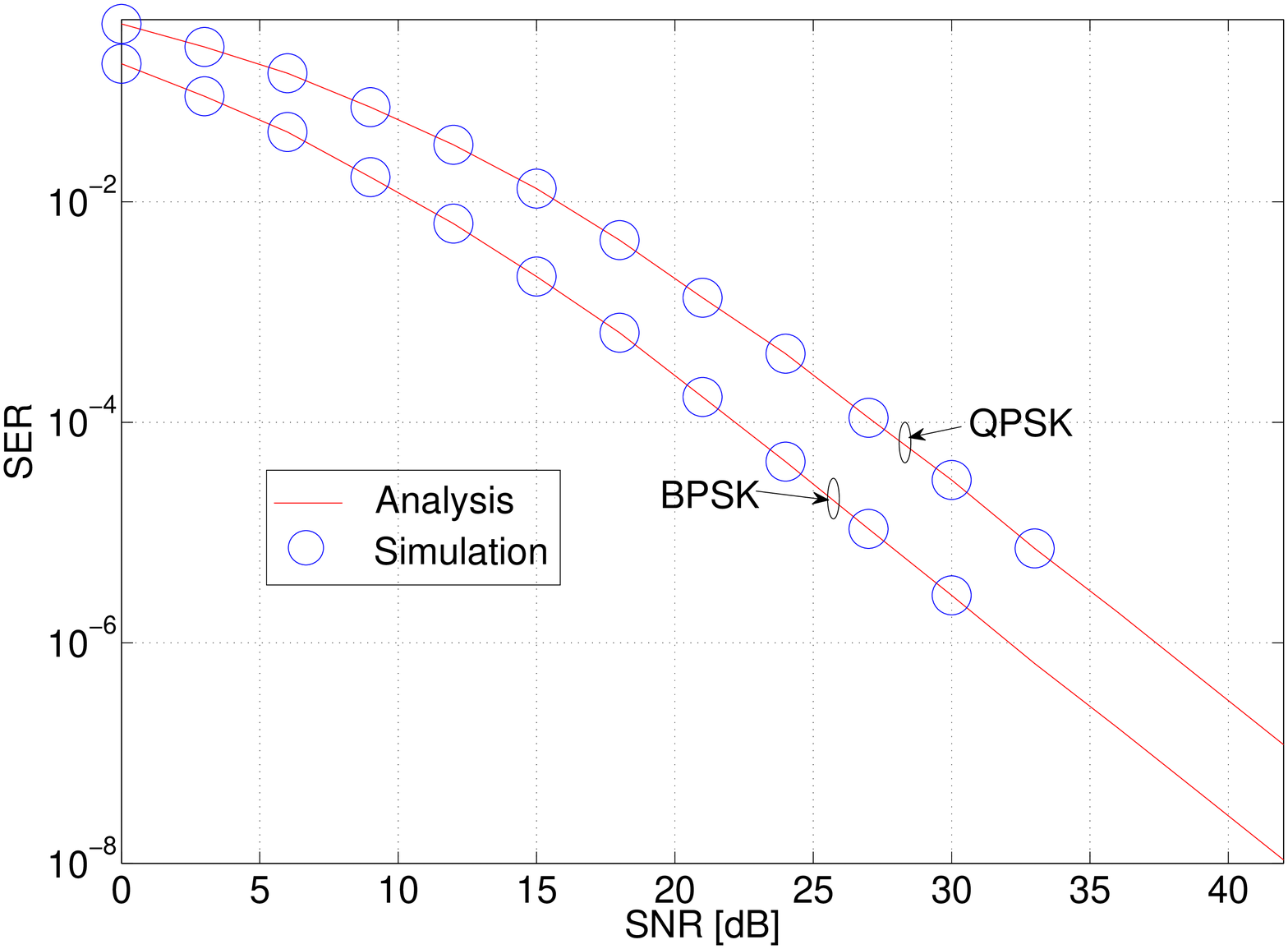,width=3.5in,height=2.5in}}\vspace*{-3mm}
\caption{Analytical and simulated BER/SER plots of the two-way AF MIMO relay system with $N_{1}=N_{2}=2$, $N_r=1$, $N_{p_1}=N_{p_2}=1$, $a=1$, Alamouti code, BPSK, and QPSK.}\vspace*{-6mm}
\label{f_10}
\end{figure}
\appendices
\vspace*{-0.05in}
\section{Proof of Theorem~\ref{the:mgf1}}
\label{app:the1}
\vspace*{-0.05in}
The m.g.f. of $\gamma^{(i)}$ can be expressed as
      \begin{align}
    &M_{\gamma^{(i)}}(s)={E}\left\{{
    \mbox{exp}(-s\gamma^{(i)})}\right\}\nonumber\\
    &={E}_{\small
    \undb{H}_{i}\undb{H}_{j}}\!\!\Bigg(\!\mbox{exp}\!\!
    \left(\!-a^2s\alpha_j\bar{\gamma}_j\mbox{Tr}{\left\{\!\undb{G}^{H}_{i,j}\!\!
   \left(Z_1\!\undb{I}_{N_i}\!\!+\!\!Z_2\undb{H}_i\!\undb{H}^{H}_i\right)^{-1}\!\!\undb{G}_{i,j}\!\right\}}\!\right)\!\!\Bigg)\!.
    \label{pa15}
    \end{align}
    Since $\undb{H}_{j}$ has a matrix
    variate complex Gaussian distribution,
    by using the results
    given in~\cite[Proof of Theorem 4.1]{LaSt:03}, we have
   \begin{align}
\!\!&M_{\gamma^{(i)}|\undb{H}_{i}}(s)\nonumber\\
&={\det\!\!\left(\undb{I}_{N_i}\!\!+\!{
a^2\!\alpha_j \bar{\gamma}_js}\undb{H}^{H}_{i}
\!\!\left(Z_1\undb{I}_{N_i}\!+\!Z_2\undb{H}_i\!\undb{H}^{H}_i\right)^{\tiny-1}
\!\undb{H}_{i}\!\right)^{-N_{j}}}\!\!\!\!\!\!\!\!\!\!\!.\label{pa17a}
    \end{align}
     Now we need to average $M_{\gamma^{(i)}|\undb{H}_{i}}(s)$ over $\undb{H}_{i}$. By performing the singular value decomposition (SVD) of $\undb{H}_{i}$, we have $\undb{H}_{i}=\undb{X}_{i}\undb{D}_{i}\undb{V}_{i}$,
    where $\undb{X}_{i}\in C^{N_i\times N_i}$ and $\undb{V}_{i}\in C^{N_r\times N_r}$ are unitary matrices, and $\undb{D}_{i}\in {\mathbb{R}}^{N_i\times N_r}$ is a diagonal matrix with singular values $l_1>l_2\dots>l_q$ ($q=\mbox{min}(N_i, N_r)$).
    From (\ref{pa17a}) and SVD factorization of $\undb{H}_{i}$, we get
   \begin{align}
\!\!&M_{\gamma^{(i)}|\undb{D}_{i}\undb{D}^T_{i}}(s)\nonumber\\&=
{\det\left(\!\undb{I}_{N_i}\!\!+\!\!{
a^2\!\alpha_j \bar{\gamma}_js}\undb{D}_{i}\undb{D}^T_{i}
\!\!\left(\!Z_1\!\undb{I}_{N_i}\!\!+\!\!Z_2\undb{D}_{i}\!\undb{D}^T_{i}\!\right)^{\!-1}
\!\right)^{-N_{j}}}\nonumber\\
&=  \prod^{q}_{j_1=1}\left({1+\frac{a^2s
    \alpha_j\bar{\gamma}_j l^2_{j_1}}{Z_1+Z_2
    l^2_{j_1}}}\right)^{-{N}_{j}}.\label{pa17b}
    \end{align}\\
Let $\lambda_1,\ldots,\lambda_q$ denote the ordered non-zero
eigenvalues of $\undb{H}^H_{i}\undb{H}_{i}$ ($\lambda_1 >
\lambda_2 > \cdots > \lambda_q$); 
we can rewrite (\ref{pa17b}) as
\beq
M_{\gamma^{(i)}|{\bf\Lambda}}(s) =
    \prod^{q}_{j_1=1}\left({1+\frac{a^2s
    \alpha_j\bar{\gamma}_j\lambda_{j_1}}{Z_1+Z_2
    \lambda_{j_1}}}\right)^{-{N}_{j}}, \,  \label{pa19}
     \eeq
       where $\lambda_{j_1}=l^2_{j_1}$ and ${\bf\Lambda}=\mbox{diag}\left(\lambda_1,\lambda_2\dots,\lambda_q\right)$. 
      The joint distribution of the eigenvalues of $\undb{H}^H_i\undb{H}_i$ can be written as \cite[Eq. (2.22)]{TuSe:04}
      \beq
f_{\bf\Lambda}\left(\lambda_1,\lambda_2\dots,\lambda_q\right)\!\!=\!\!\kappa^{-1}\prod^{q}_{j_1<j_2}\left(\lambda_{j_2}\!\!-\!\!\lambda_{j_1}\right)^2
      \prod^{q}_{k=1}\lambda^{p-q}_{k}\mbox{exp}(-\lambda_{k}), \label{pa19a}
      \eeq
      where $\lambda_1 >
\lambda_2 > \cdots > \lambda_q>0$. 
      By averaging
      (\ref{pa19}) over ${\bf\Lambda}$, with the help of 
      (\ref{pa19a}) and \cite[Corollary 2]{ChWiZa:03}, we have
     \beq
      M_{\gamma^{(i)}}(s)=\kappa^{-1}\det\left(\undb{J}(s)\right),
      \label{pa200}
      \eeq
      where $\kappa=\prod_{l=1}^{q}\Gamma{(p-l+1)}
      \Gamma{(q-l+1)}$ 
      and $\undb{J}(s)$ is
    a $q \times q$ Hankel matrix with
    $(t,v)$-th element
      \beq
      \!\!{{{J}}_{tv}}(s)\!\!=\!\!
      \int^{\infty}_{0}\!\!{\lambda^{\nu_{tv}-1}\!\!
    \left(\frac{\left(1+
     \frac{Z_2}{Z_1}\lambda\right)}{1+
    \frac{Z_2}{Z_1}\lambda\left(1
           +\frac{a^2s\alpha_j\bar{\gamma}_j}{Z_2}\right)
      \lambda}\!\!\right)^{N_{j}}\!\!\!\!\!\exp{\!(-\!\lambda)}
      d\lambda}.\label{pa2111}
      \eeq
After solving the integral in (\ref{pa2111}) by
using~\cite[Eq.~(3.383.5)]{GrRy:03},
we obtain (\ref{pa22}).
\begin{figure}[t!]
\vspace*{-5mm}
\vspace*{-3mm}
\centerline{\psfig{file=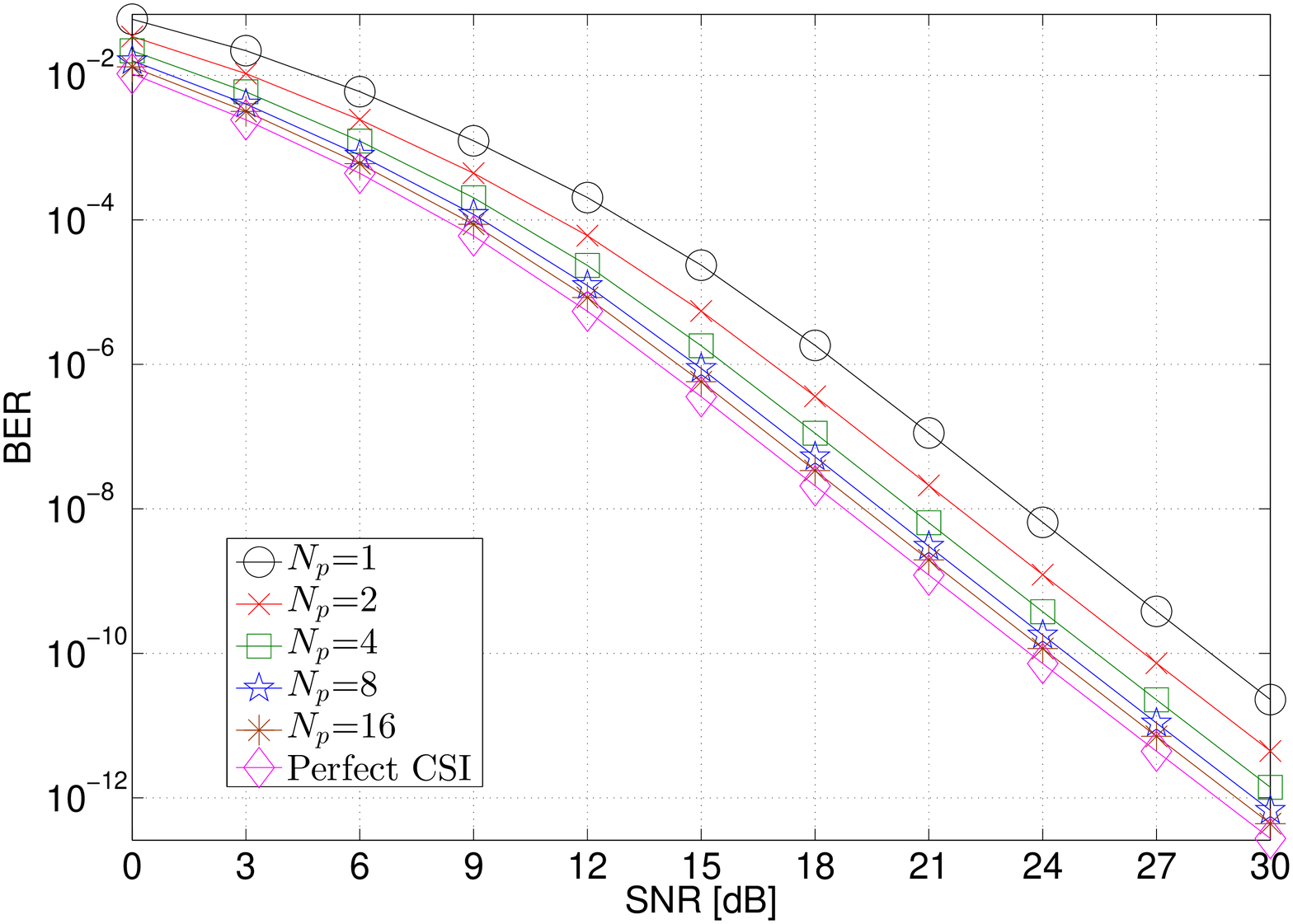,width=3.5in,height=2.45in}}\vspace*{-3mm}
\caption{Analytical BER plots of the two-way AF MIMO relay system with $N_{1}=N_{2}=N_r=2$, $N_p=N_{p_1}=N_{p_2}=1,2,4,8,16$, $a=1$, Alamouti code, and BPSK constellation.}\vspace*{-6mm}
\label{f_7}
\end{figure}
\vspace*{-0.2in}
\section{Proof of Theorem 2}
\label{dec_div}
It can be seen from~\cite[Eqs. (13.5.6), (13.5.9), and (13.5.12)]{AbSt:70} that
the confluent hypergeometric function $U(a_1,b_1,z)$
of the second kind can be asymptotically
approximated for $z \ll 1$ (small $z$) as\vspace*{0.1in} 
 \begin{align}
U(a_1,b_1,z)\approx\left\{\begin{array}{cc}
\frac{\Gamma(b_1-1)}{\Gamma{(a_1)}}z^{1-b_1} \, ,
&  \mbox{if} \quad b_1\geq 2,\\
          -\frac{1}{\Gamma{(a_1)}}(
          \ln{z}+\Psi(a_1)) \, ,
          &  \mbox{if} \quad b_1=1,  \\
          \frac{\Gamma{(1-b_1)}}{
          \Gamma{(1+a_1-b_1)}} \, ,
          & \mbox{if} \quad b_1\leq 0,\end{array}\right. \label{d1}
          \end{align}\\
 where $\Psi(\cdot)$ is the digamma
 function~\cite{AbSt:70} and $a_1>0$.
 We consider the case when
 $\mbox{min}(N_{r},N_{i})= 1$ and $p<N_{j}$.
 In this case, the m.g.f.
 of the received SNR of the ML decoder can be
 expressed using~(\ref{pa11}) as
 \beq
M_{\gamma^{(i)}}(s)\!=\!\frac{{\sum^{p-1}_{k=0}{\phi(k,s)}}
\!+\!\phi(p,s)\!+\!\sum^{N_{j}}_{k=p+1}{\phi(k,s)}}
{{\left(1
+\frac{a^2s\alpha_j \bar{\gamma}_j}{{Z_2}}
\right)^{p+N_{j}}}\left(\frac{Z_2}{Z_1}\right)^{p}} \, ,
\label{d2}
 \eeq
 where
 $\phi(k,s)\!\!=\!\!({^{N_{j}}_{k}})\!\!
 \left(\!\!\frac{a^2s\alpha_j \bar{\gamma}_j}{{Z_2}}\right)^{k}
 \!\!\Gamma(p) U\!\!\left(\!\!p, \, p\!\!+\!\!1\!\!-\!\!k,\! \frac{Z_1Z_2^{-1}}
 {(1+\frac{a^2\alpha_j s \bar{\gamma}_j}{Z_2})}\!\!\right)\!\!. \vspace*{0.1in}
 $
 Let us assume that $\bar{\gamma_1}=\bar{\gamma_2}=\bar{\gamma}$ (all links are identically distributed) and
 $b=\beta\bar{\gamma}$, $\beta\in{\mathbb{R}}^+$, 
 for diversity calculations.
 It can be shown from (\ref{a}) that $a^2={\beta}/{\left(2N_rT\right)}$, for very large values of $\bar{\gamma}$.
 Therefore, for $|s|\bar{\gamma} \gg 1$ (large $s \bar{\gamma}$),
from (\ref{d1}) and (\ref{d2}), we have
 \begin{align}
 \sum^{p-1}_{k=0}{\phi(k,s)}&\approx
 \left(\frac{a^2\alpha_j \bar{\gamma} s}{Z_2}\right)^{p}
 \sum^{p-1}_{k=0}({^{N_{j}}_{k}})
 \Gamma{(p-k)} \, ,\nonumber\\
 \!\!\!\!\!\!\!\!\!\!\hspace*{-5mm}\phi(p,s)\approx&\left(\!\frac{a^2\alpha_j \bar{\gamma} s}{Z_2}\right)^{p}\!\!\!\!({^{N_{j}}_{p}})
 \bigg(\!\!\!\!-\!\!\Psi(p)\!+\!\ln \alpha_j \!+\!\ln s \!+\! \ln \frac{a^2\bar{\gamma}}{Z_2}\bigg),\nonumber\\
  \sum^{N_{j}}_{k=p+1}{\phi(k,s)}&\approx
  \sum^{N_{j}}_{k=p+1}\left(\frac{a^2\alpha_j \bar{\gamma} s}{Z_2}\right)^{k}({^{N_{j}}_{k}})
  \frac{\Gamma{(k-p)}\Gamma{(p)}}{\Gamma{(k)}} \, ,
  \label{d5}
  \end{align}
 where $\alpha_j$, $a$, $Z_1$, and $Z_2$ are independent of $s$ and $\bar{\gamma}$. 
Taking the most dominant power of
    $s \bar{\gamma}$ in (\ref{d5}), 
    the m.g.f., given in (\ref{d2}), can be approximated for
    large $s \bar{\gamma}$ as
  $
  M_{\gamma^{(i)}}(s)\approx \frac{\eta_{1}}{(s\bar{\gamma})^{p}} \, ,
  $
 where $\eta_{1}$
 is a constant independent of $s$ and $\bar{\gamma}$.
 Hence, the diversity order
 of the ML decoder is $p$ for
 $\min(N_r,N_i)=1$ and
 $p<N_{j}$. For $\min(N_r,N_i)=1$ and $p \geq N_{j}$,  we can obtain the approximate m.g.f. by
 a method similar to that for $p < N_j$ as
$M_{\gamma^{(i)}}(s)\approx {\eta_{2}}/{(s\bar{\gamma})^{N_j}}$,
where $\eta_{2}$
is a constant independent of $s$ and $\bar{\gamma}$,
resulting in a diversity order of $N_j$.
  

\vspace*{-0.1in}
\begin{thebibliography}{07}
      \bibitem{GuGoPo:08}
D. Gunduz, A. Goldsmith, and H. V. Poor,
``MIMO two-way relay
channel: Diversity-multiplexing trade-off analysis,'' In \emph{Proc. 42nd Asilomar
Conference on Signals, Systems and Computers}, Pacific Grove, CA, Oct.
2008, pp. 1474 - 1478. 

\bibitem{WyOeBj:08}
R. F. Wyrembelski, T. J. Oechtering, I. Bjelakovic, C. Schnurr, and H.
Boche,
``Capacity of Gaussian MIMO bidirectional broadcast
channels,'' In \emph{Proc. IEEE International Symposium on Information Theory (ISIT 2008)},
Toronto, Canada, July 6-11, 2008, pp. 584 - 588.

\bibitem{ArRtRk:12}
Arti M.K., R. K. Mallik, and R. Schober,
``Beamforming and combining
in two-way AF MIMO relay networks,'' \emph{IEEE Comm. Lett.}, vol. 17,
no. 7, pp. 1400-1403, July 2013.

\bibitem{ArRtRk:13}
Arti M.K., R. K. Mallik, and R. Schober,
``Channel estimation and decoding of
OSTBC in two-way AF MIMO relay networks,'' In \emph{Proc. IEEE Vehicular Technology
Conference (VTC) - Fall},
Las Vegas, USA, 2-5 Sept. 2013.

\bibitem{LiHo:10}
 L. Song, G. Hong, B. Jiao, and M. Debbah,
``Joint relay selection and analog network coding using differential modulation in two-way relay 
channels,'' \emph {IEEE \mbox{Tr}ans. Veh. Techn.}, vol. 59,
no. 6, pp. 2932--2939, July 2010.
\bibitem{GaZhLi:09}
F. Gao, R. Zhang, and Y.-C. Liang,
``Optimal channel estimation and
training design for two-way relay networks,'' \emph{IEEE \mbox{Tr}ans. Commun.},
vol. 57, no. 10, pp. 3024-3033, Oct. 2009. 
\bibitem{FaSh:11}
Z. Fang and H. Shan ``Comparison of channel estimation schemes for
MIMO two-way relaying systems,'' In
\emph{Proc. Cross Strait Quad-Regional
Radio Science and Wireless Technology Conference (CSQRWC)},
Ningbo, China, 
 July 26-30, 2011, pp. 719 - 722. 
 \bibitem{AlRo:10}
A. Y. Panah and R. W. Heath,
``MIMO two-way amplify-and-forward relaying
with imperfect receiver CSI,"
\emph {IEEE \mbox{Tr}ans. Veh. Techn.},
 vol. 59, no. 9, pp. 4377-4387, Nov. 2010.
 
 \bibitem{JuHa:12}
J. Zou, H. Luo, M. Tao,
and R. Wang,
``Joint source and relay optimization for
non-regenerative MIMO two-way relay systems with imperfect CSI,"
\emph{IEEE \mbox{Tr}ans. Wireless Commun.},
 vol. 11, no. 9, pp. 3305-3315, Sept. 2012. 
  \bibitem{RaRo:11}
 R. M. Legnain and R. H. M. Hafez, ``MIMO two-way relay communication based on Alamouti code,''
 \emph{in IEEE Canadian Conference on Electrical and Computer Engineering (CCECE)},
 Niagra Falls, Canada, May 8-11, 2011,
 pp. 1-5.
 \bibitem{RaRoIa:12}
 R. M. Legnain, R. H. M. Hafez, I. D. Marsland, and A. M. Legnain ``Two-way quantize-and-forward relaying with STBC,''
 \emph{in IEEE Canadian Conference on Electrical and Computer Engineering (CCECE)},
 Canada, Apr. 29-May 02, 2012,
 pp.~1-5.
   \bibitem{LaSt:03}
E. G. Larsson and P. Stoica, \emph{Space-Time Block Coding for Wireless
Communications.} Cambridge, UK: Cambridge University Press, 2003.
 \bibitem{Manav:07}
M. R. Bhatnagar and A. Hj{\o}rungnes, ``SER expressions for double
differential modulation,''
in \emph{Proc. IEEE Information Theory Workshop (ITW 2007)},
Bergen, Norway, July 2007, pp. 1-5.
\bibitem{Manav:08}
M. R. Bhatnagar, A. Hj{\o}rungnes, and L. Song, ``Cooperative communications over flat fading channels with carrier offsets: A double-differential modulation Approach,'' \emph{EURASIP Journal on Advances in Signal Processing,} vol. 2008, Article ID 531786,
pp. 1-11, April 2008.
\bibitem{Manav:13}
M. R. Bhatnagar and O. Tirkkonen, ``PL decoding in double differential modulation based decode-and-forward cooperative system,''
\emph{ IEEE Commun. Letters,} vol. 17, no. 5, pp. 860-863, May 2013.
 \bibitem{ChWiZa:03}
H. Shin and J. H. Lee, ``Performance analysis of space-time
block codes over keyhole Nakagami-$m$ fading channels,"
\emph {IEEE \mbox{Tr}ans. Veh. Technol.}, vol. 53, no. 2,
pp. 351--362, Mar. 2004.
\bibitem{SiAl:05}
M. K. Simon and M.-S. Alouini, {\it
Digital Communication Over
Fading Channels}, 2nd ed. Hoboken, NJ: Wiley, 2005.
\bibitem{GrRy:03}
I. S. Gradshteyn and I. M. Ryzhik, {\it
Tables of Integrals,
Series and Products}, 6th ed. San Diego, CA,
USA: Academic Press, 2000.
\bibitem{LeHaMa:12}
R. M. Legnain, R. H. M. Hafez, and I. D. Marsland,
``BER analysis of three-phase XOR-and-forward relaying
using Alamouti STBC", \emph{IEEE Commun. Lett.}, vol. 16, no. 9, pp.
1458--1461, Sept. 2012.
\bibitem {TuSe:04}
 A. M. Tulino and S. Verd{\'u},  ``Random matrix theory and wireless
communications,''\emph{
Foundations Trends Commun. Inf. Theory},
vol. 1,
no. 1, pp. 1-163, 2004.

\bibitem{AbSt:70}
M. Abramowitz and I. A. Stegun, {\it Handbook of
Mathematical Functions.} New York,
USA: Dover Publications, Inc., 1972.
\end{thebibliography}
\end{document}